\documentclass[twocolumn,pra,aps,superscriptaddress]{revtex4}
\usepackage{epsfig}
\usepackage{amsbsy}
\usepackage{amsmath}
\usepackage{amsfonts}
\usepackage{graphicx}
\usepackage{latexsym}
\usepackage{xcolor}

\providecommand{\U}[1]{\protect\rule{.1in}{.1in}}

\begin{document}

\title[Short title for running header]
{Exact dynamics and bound states of a cavity coupled to a two-dimensional reservoir}

\author{Heng-Na Xiong}
%\email{hnxiong@zjut.edu.cn}
\affiliation{Department of Applied Physics, Zhejiang University of Technology,
Hangzhou 310023, P. R. China}

\author{Da-Wei Ye}
\affiliation{Department of Applied Physics, Zhejiang University of Technology,
Hangzhou 310023, P. R. China}

\author{Yang Yang}
\affiliation{Department of Applied Physics, Zhejiang University of Technology,
Hangzhou 310023, P. R. China}

\author{Hongli Zhu}
\affiliation{The Zhejiang Laboratory,
Hangzhou 310023, P. R. China}

\author{Yixiao Huang}
\affiliation{School of Science, Zhejiang University of Science and Technology, Hangzhou, Zhejiang, 310023, China}

\author{Stefano Longhi}
\email{stefano.longhi@polimi.it}
\affiliation{Dipartimento di Fisica, Politecnico di Milano, Piazza L. da Vinci 32, I-20133 Milano, Italy}
\affiliation{IFISC (UIB-CSIC), Instituto de Fisica Interdisciplinary Sistemas Complejos- Palma de Mallorca, Spain}

\author{Fanxin Liu}
\email{liufanxin@zjut.edu.cn}
\affiliation{Department of Applied Physics, Zhejiang University of Technology,
Hangzhou 310023, P. R. China}

\date{\today}

\begin{abstract}

We demonstrate a robust scheme for quantum information storage based on bound states in a two-dimensional coupled-cavity array. When a target cavity is tuned to resonance with the array, a bound state in the continuum (BIC) emerges, coexisting with two conventional bound states outside the band. The resulting dynamics reflects a delicate interplay between these bound states, which can be fully captured through exact analytical solutions. In the weak-coupling regime, the BIC dominates, enabling perfect and persistent information storage. At stronger coupling, all bound states contribute, leading to oscillatory behavior and reduced storage fidelity. These results, valid at both zero and finite reservoir temperatures and further supported by a single-particle framework, reveal distinctive non-Markovian features in continuous-variable systems and highlight the potential of photonic lattices for scalable all-optical decoherence-free quantum memory platforms.

\end{abstract}

\maketitle

\section{Introduction}

Decoherence induced by the environment remains a key challenge for the development of quantum information and computation technologies \cite{Nielson00}. In the weak system-environment coupling regime, open quantum systems typically exhibit Markovian dynamics, where quantum information decays exponentially over time. However, with stronger coupling, non-Markovian memory effects become prominent, enabling partial preservation of quantum coherence and information \cite{Rivas14,Breuer16,Vega17,Li18,Zhang19,Carmele19}.

While natural thermal environments (e.g., those with Ohmic-type spectra \cite{{Hu92,Zhang12,Xiong10}}) are commonly considered, structured reservoirs offer enhanced control and richer dynamics \cite{Dong09,Wu10,Tan11,Lei12,Hoeppe12,Bradford13,Tufarelli13,Tufarelli14,Goban14,Xiong15,Lo15,Roy17,Xiong17,Xiong18,Shen19,
Shen19b,Sundaresan19,Andersson19,Guo20,Burgess21,Ferreira21,Du21,Scigliuzzo22,Zhu22,Xiong22}. Among them, coupled cavity arrays (CCAs) stand out for their experimental feasibility in photonic-crystal \cite{Liu05,Faraon,Saxena23}, semiconductor \cite{Bogaerts12,Zhang2019}, and other cavity systems \cite{Vetsch10,Yamamoto08,Elshaari12,Akimov07}. CCAs provide versatile platforms for quantum information processing \cite{Zheng13,Paulisch16} and quantum simulation \cite{Carusotto13,Noh17,Sheremet23}, offering tunable system-reservoir and inter-cavity coupling \cite{Liu05,Faraon,Saxena23}. A key feature of CCAs is their band structure, which can support bound states outside the continuum (BOCs) residing within band gaps. These states are essential for achieving non-Markovian and dissipationless dynamics \cite {Wu10,Tan11,Lei12,Lo15,Xiong17,Xiong18,Guo20,Scigliuzzo22,Xiong22}. However, BOCs are evanescent and require high-index materials to open complete gaps, limiting practical implementation \cite{Joannopoulos2008,Sakoda2005}.

Recently, bound states in the continuum (BICs) have emerged as a compelling alternative \cite{Hsu2016,Limonov2017,Koshelev2019,Azzam2021,Joseph2021,Zhong2023,Kang2023}. Unlike BOCs, BICs exist within the propagating spectrum yet remain decoupled from the environment, offering non-radiative properties, strong localization, and enhanced design flexibility with lower-index materials. BICs have been realized in various photonic structures and enable applications in lasing, filtering, sensing, imaging, nonlinear optics, and entangled-photon generation.

Despite this progress, their use for quantum information storage has been under explored. As decoherence-free (dark) states, BICs hold great promise in this context. Recent works suggest their potential in topological quantum batteries \cite{Lu25} and entanglement filtering in optomechanics and photonics \cite{Shang24,Selim25,Longhi25}. In \cite{Lim2023}, BICs were shown to outperform BOCs for information storage in a giant-atom model within a 1D non-Markovian photonic lattice. Two-dimensional (2D) CCAs \cite{Altug04,Alija06,Gourdon12,Khanikaev17,Little00,Ohno22,Yang18} further enrich the landscape for quantum technologies \cite{Majumdar12,Xu13,Barik18,Chen21}. Prior studies \cite{Tudela17L,Tudela17A} uncovered non-perturbative dynamics in 2D CCAs arising from singularities in the density of states, but focused on emitter-based systems and did not identify BICs.

%In this work, we study the exact dynamics of a continuous-variable system (a target cavity) coupled to a 2D CCA. We demonstrate that when the cavity is resonant with the array, a BIC emerges alongside two BOCs. This leads to distinctive, analytically tractable dynamics and enables perfect quantum information storage even in the weak coupling regime, highlighting a novel mechanism for \textbf{robust quantum memory: change it as decoherence-free quantum memory???}.

In this work, we study the exact dynamics of a continuous-variable system, specifically a target cavity, coupled to a two-dimensional coupled-cavity array. We demonstrate that when the cavity is tuned to resonance with the array, a BIC emerges alongside two conventional BOCs. This leads to distinctive, analytically tractable dynamics and enables nearly perfect preservation of quantum information in the weak-coupling regime. The BIC acts as a decoherence-free state, effectively trapping information within the target cavity and preventing leakage into the reservoir. This dissipationless behavior is universal for arbitrary initial states, including single-photon, multi-photon, and superposition states.

As a potential application, this all-light photonic setup can function as a decoherence-free quantum memory \cite{Lvovsky2009,Afzelius2015,Heshami2016,MaL2020,WangYF2023,Lukin2001,Phillips2001,Duan2001,RRR2}, where the information remains entirely encoded in the light \cite{Lvovsky2009,RRR3, RRR4,RRR5} and is delayed or stored via the BIC, without relying on atomic or matter-based degrees of freedom. More broadly, our results illustrate how engineered BICs in photonic lattices enable precise control over non-Markovian dynamics and light-matter interactions in a scalable, chip-integrated platform.

The paper is structured as follows: Sec. II presents the system Hamiltonian. In Sec. III, we analyze the exact dynamics and identify the bound states at both zero and finite reservoir temperatures. Sec. IV explores information distribution via a single-particle framework. Conclusions and outlook for future research directions follow in Sec. V.

\begin{figure}[ptb]
	\centering
	\includegraphics[width=8cm]{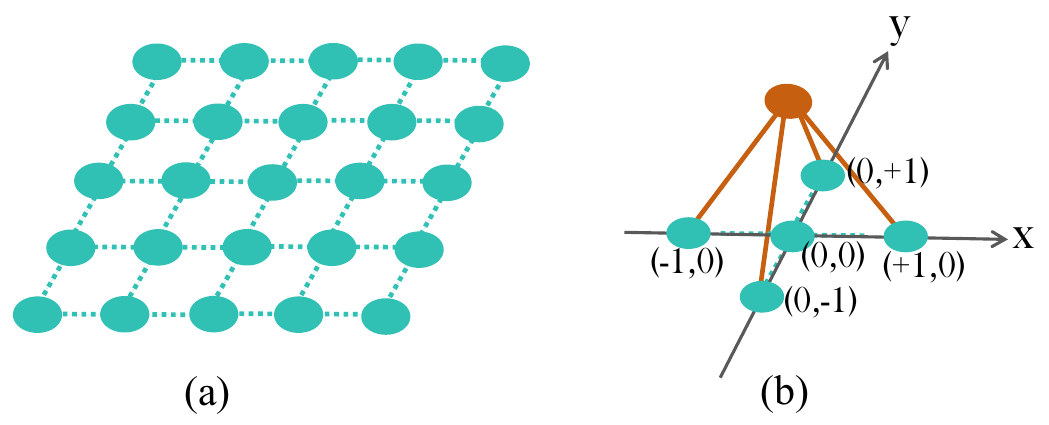}\newline
	\caption{\textbf{Schematic diagram of the model.} (a) The 2D CCA (denoted by green circles) with tight-binding interactions (denoted by the green dotted lines). (b) The target cavity (denoted by a brown circle) located directly above the center $(0,0)$-th cavity of the 2D CCA and its interactions with the 2D CCA (denoted by four brown lines).}
	\label{fig-2D}
\end{figure}

\section{Hamiltonian}
We consider a continuous-variable quantum system of a target cavity interacting with a 2D $(2N+1)\times (2N+1)$ square CCA, as shown in Fig. \ref{fig-2D}. The Hamiltonian of the total system reads (we assume $\hbar=1$)
\begin{equation}\label{HT}
	{{H}}={{H}_{0}}+{{H}_{R}}+{{H}_{I}},
\end{equation}
with
\begin{eqnarray}
{{H}_{0}}&=&{{\omega}_{c}}{a}^{\dagger}{a}, \nonumber\\
{{H}_{R}}&=&\sum\limits_{{\vec{r}}}{{{\omega}_{0}}b_{{\vec{r}}}^{\dagger}{{b}_{{\vec{r}}}}}-
\sum\limits_{\left\langle\vec{r},\vec{r}'\right\rangle}{{\xi}_{0}}{\left(b_{{\vec{r}}}^{\dagger}{{b}_{{\vec{r}'}}}
+b_{{\vec{r}'}}^{\dagger}{{b}_{{\vec{r}}}}\right)} , \nonumber\\
{{H}_{I}}&=&\sum_{\vec{s}}\xi({ b_{{{{\vec{s}}}}}^{\dagger}{a}+a^{\dagger }{{b}_{{{{\vec{s}}}}}} )}.
\end{eqnarray}
Here $H_0$ is the Hamiltonian of the target cavity located directly above the center ${{\vec{r}}_{0}}=\left({{{{0}}}},{{{{0}}}}\right)$ of the 2D CCA,
with $a^{\dagger}$ and $a$ being the creation and annihilation operators of the cavity mode, respectively. The frequency of the target cavity is denoted by ${{\omega}_{c}}$. The second term $H_R$ is the Hamiltonian of the reservoir. $b_{{\vec{r}}}^{\dagger}$ and ${{b}_{{\vec{r}}}}$ are the creation and annihilation operators of the cavity mode at position $\vec{r}=(x,y)$ with frequency $\omega_0$ . Here the summations over $x$ and $y$ range from $-N$ to $+N$. Moreover, the summation over ${\left\langle\vec{r},\vec{r}'\right\rangle}$ only accounts for the nearest-neighbor coupling (denoted by the grey dotted lines in Fig.~\ref{fig-2D}(a)) between the cavities, with all coupling strengths equal to ${{\xi}_{0}}$. The last term
$H_I$ describes the interactions between the target cavity and the four nearest-neighboring cavities of the center cavity at the four positions ${{\vec{s}}}= \{\left(\pm1,0 \right), \left( 0,\pm 1\right)\}$ with the coupling strength $\xi$ (denoted by the four brown lines in Fig.~\ref{fig-2D}(b)). We note that our geometrical configuration differs from the more conventional setups, where emitters couple only to the cavity in which they reside~\cite{Tudela17L,Tudela17A}. In our model, we neglect the interaction between the target cavity and both the central cavity at \((0,0)\) and the next-nearest-neighboring cavities at \((\pm1, \pm1)\) in the CCA. This approximation is physically justified, as demonstrated in Appendix~\ref{app-phys-model} using different ring cavity configurations.

 Analogous to the 1D case \cite{Xiong17,Xiong18}, by assuming $N \rightarrow \infty$, we can perform a 2D Fourier transformation  ${{b}_{{\vec{k}}}}=\sum_{\vec{r}}{{{e}^{-i\vec{k}\cdot \vec{r}}}{{b}_{{\vec{r}}}}}$ and $b_{\vec{r}}=\iint \frac{d{{\vec{k}}}}{(2\pi)^2}{{e}^{i\vec{k}\cdot \vec{r}}}b_{\vec{k}}$ with $\vec{k}=(k_x,k_y)$ the momentum vector. In this way, we diagonalize the reservoir Hamiltonian as follows
\begin{equation}\label{HR-k}
{{H}_{R}}=\iint \frac{d{{\vec{k}}}}{(2\pi)^2}~{{{\omega }_{{\vec{k}}}}b_{{\vec{k}}}^{\dagger }{{b}_{{\vec{k}}}}}.
\end{equation}
 Here, the dispersion relation of the 2D CCA is \cite{Tudela17L,Tudela17A,Economou06}
\begin{equation}\label{Dispersion}
  {{\omega }_{{\vec{k}}}}={{\omega }_{0}}-2{{\xi}_{0}}\left( \cos {{k}_{x}}+\cos {{k}_{y}} \right),
\end{equation}
 which is similar to the 1D case. The energy band spans the range from $\omega_0-4\xi_0$ to $\omega_0+4\xi_0$. Consequently, the interaction between the target cavity and the reservoir becomes
 \begin{equation} \label{HI-k}
{{H}_{I}}=\iint \frac{d{{\vec{k}}}}{(2\pi)^2}~{\left( {{V}_{{\vec{k}}}}b_{{\vec{k}}}^{\dagger }a+V_{{\vec{k}}}^{*}{{b}_{{\vec{k}}}}a^{\dagger } \right)},
 \end{equation}
 where the coupling strength is
 \begin{equation}\label{eq-Vk}
{{V}_{{\vec{k}}}}
 ={2\xi }{{e}^{i\vec{k}\cdot {{{\vec{r}}}_{0}}}}\left( \cos {{k}_{x}}+\cos {{k}_{y}} \right)={\eta }{{e}^{i\vec{k}\cdot {{{\vec{r}}}_{0}}}}\left( {{\omega }_{0}}-{{\omega }_{{\vec{k}}}} \right),
 \end{equation}
with $\eta \equiv \frac{\xi }{\xi _{0}}$ the dimensionless coupling strength. Here the coupling ${V}_{{\vec{k}}}$ is proportional to the frequency ${\omega }_{{\vec{k}}}$. Notably, $V_{\vec{k}}$ can vanish when $\omega_{\vec{k}}=\omega_0$, suggesting the presence of a BIC inside the band, as will further explored below.

Notably, although the Hamiltonian (\ref{HT}) omits the coupling between the target cavity and the central cavity at (0,0), our model can be readily extended to include it, thereby enhancing its physical realizability. As demonstrated in Appendix \ref{app-phys-model}, this additional coupling merely introduces a constant shift to $V_k$ and a frequency shift to the BIC condition, as shown in Eqs. (\ref{eq-Vkk})-(\ref{eq-eta00}). The strategic omission of this term in our core framework serves to clearly isolate the fundamental interaction mechanism responsible for the BIC formation. That is, the BIC originates dominantly from the four nearest-neighbor interaction topology.

\section{Exact Dynamics of the System}
In this section, we first introduce the exact master equation of the system to address its dynamical properties. We then reformulate the conditions for the existence of bound states of the total system within our notation. Based on this framework, we numerically and analytically investigate the exact dynamics of the system in the presence of an initially zero-temperature reservoir, explicitly highlighting the emergence of bound states, including both BOCs and BICs. Finally, we extend our discussion to the exact dynamics of the system in the case of a finite-temperature reservoir.
\subsection{Exact Master Equation and Bound States}

\begin{figure}[ptb]
\centering
\includegraphics[width=8cm]{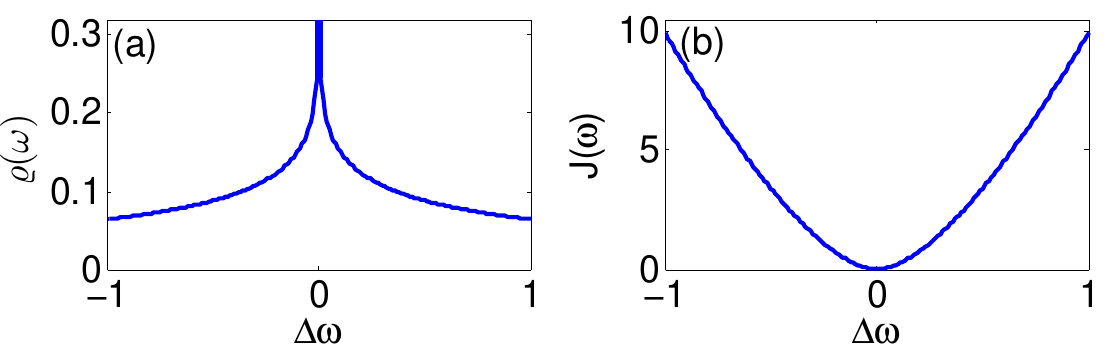}\newline
\caption{(a) Density of state $\varrho(\omega)$ and (b) spectral density $J(\omega)$ for a target cavity coupled to 2D CCAs. Here we set $\eta=1$. }
\label{fig-Jw}
\end{figure}

For the continuous-variable quantum system described above, the exact master equation governing the relaxation dynamics of the quantum field in the target cavity is given by~\cite{Hu92,Xiong10,Zhang12,Xiong15}:
\begin{align}\label{Master equation}
	\frac{\text{d}}{\text{d}t}\rho (t)=
	& -i{{{\omega }'}_{c}}\left( t \right)\left[ {{a}^{\dagger }}a,\rho (t) \right]+ \nonumber\\
	& \kappa (t)\left\{ 2a\rho (t){{a}^{\dagger }}-{{a}^{\dagger }}a\rho (t)-\rho (t){{a}^{\dagger }}a \right\} + \nonumber\\
	& \tilde{ \kappa}(t)\left\{ {{a}^{\dagger }}\rho (t)a + a\rho (t){{a}^{\dagger }}-{{a}^{\dagger }}a\rho (t)-\rho (t)a{{a}^{\dagger }} \right\}.
\end{align}
This equation is derived using the Feynman-Vernon influence functional formalism within the coherent-state path integral framework. It assumes that the system and the reservoir are initially uncorrelated, and that the reservoir is initially in a thermal state.
In Eq.~\eqref{Master equation}, the coefficient \(\omega_c'(t)\) represents the renormalized frequency of the system, while \(\kappa(t)\) and \(\tilde{\kappa}(t)\) denote the dissipation and fluctuation coefficients of the thermal reservoir, respectively. These coefficients are determined by the following expressions:
\begin{align}\label{coefficients}
	{{{\omega }'}_{c}}(t) & =-\text{Im}|\dot{u}(t){{u}^{-1}}(t)|, \nonumber\\
	\kappa (t) & =-\text{Re}|\dot{u}(t){{u}^{-1}}(t)|, \nonumber\\
	\tilde{ \kappa}(t) & = \dot{v}(t) - 2v(t)\text{Re}|\dot{u}(t){{u}^{-1}}(t)|,
\end{align}
with Green  functions $u(t)$ and $v(t)$ satisfying the equations
\begin{align}\label{eq-ut-vt}
	\dot{u}(t)&+i{{\omega }_{c}}u(t )+ \int_{{0}}^{t }{\text{d}}{\tau } g(t -{\tau })u({\tau })=0, \nonumber\\
	v(t)=& \int_{0}^t d\tau_1 \int_{0}^t d\tau_2 u(t-\tau_1) \tilde{g}(\tau_1-\tau_2) u^*(t-\tau_2),
\end{align}
where the initial conditions are $u(0)=1$ and $v(0)=0$. The integral kernels $g(\tau)$ and $\tilde{g}(\tau)$ are defined as
\begin{eqnarray}
  g(\tau) &=& \int^{\omega_0+4\xi_0}_{\omega_0-4\xi_0} \frac{d\omega}{2\pi} J(\omega) e^{-i\omega\tau}, \nonumber\\
  \tilde{g}(\tau) &=& \int^{\omega_0+4\xi_0}_{\omega_0-4\xi_0} \frac{d\omega}{2\pi} J(\omega) \bar{n}(\omega,T) e^{-i\omega\tau}.
\end{eqnarray}
The kernels characterize the non-Markovian memory effects of the reservoir on the system and encapsulate the complete information about the dissipation and thermal fluctuation dynamics induced by the reservoir. The function \( J(\omega) \) denotes the spectral density of the reservoir, which describes the frequency-dependent distribution of the system-reservoir coupling strength (see Eq.~\eqref{Jw}). The quantity \( \bar{n}(\omega, T) = \frac{1}{e^{\omega/(k_B T)} - 1} \) represents the initial average photon number distribution of the thermal reservoir.
At absolute zero temperature (\( T = 0 \)), the reservoir is in the vacuum state, and the Green's function \( v(t) \) vanishes. In general, different spectral densities lead to distinct solutions for the Green's functions \( u(t) \) and \( v(t) \), thereby giving rise to a variety of system dynamics.

To obtain the spectral density for the 2D reservoir, we need to calculate the density of states $\varrho\left(\omega\right)=\iint{\frac{d\vec{k}}{{{\left(2\pi\right)}^{2}}}~\delta\left(\omega-{{\omega}_{{\vec{k}}}}\right)}$ first \cite{Economou06}
\begin{equation}\label{Dw}
	\varrho\left( \omega  \right)=\frac{1}{2{{\pi }^{2}}{{\xi }_{0}}}K\left( \sqrt{1-({\Delta\omega})^{2}} \right)
\end{equation}
with $\Delta\omega\equiv\frac{\omega -\omega_0}{4\xi_0}$.
The function $K(x)$ is the complete elliptic integral of the first kind, $K\left( x \right)=\frac{1}{2}\int_{0}^{\pi }{\frac{d\alpha }{\sqrt{1-{{x}^{2}}{{\cos }^{2}}\alpha }}}$.
Then we have the spectral density $J\left( \omega  \right)=2\pi \varrho\left( \omega  \right) {{\left| V\left( \omega  \right) \right|}^{2}}$  as
\begin{align}\label{Jw}
	J(\omega)=\frac{{16\xi_0{\eta }^{2}}}{{{\pi }}} {{\left( \Delta\omega  \right)}^{2}}K\left( \sqrt{1-{{\left( \Delta\omega\right)}^{2}}} \right)
\end{align}
with $\Delta\omega \in (-1,1)$ indicating the spectral band $(\omega_0-4\xi_0,\omega_0+4\xi_0)$.

As detailed in the Appendix \ref{app-BS}, the condition for the existence of a bound state of the total system with eigenenergy $\Omega$ is \cite{Miyamoto2005,Longhi2007}
\begin{eqnarray}\label{eq-bd-cond-1}
\Omega-\omega_c-\Sigma(\Omega)=0,
\end{eqnarray}
where $\Sigma(\Omega)=\int^{\omega_0+4\xi_0}_{\omega_0-4\xi_0} \frac{d\omega}{2\pi} \frac{J(\omega)}{\Omega-\omega}$.
When the frequency $\Omega$ is outside the continuum (band) $({\omega_0+4\xi_0},{\omega_0-4\xi_0})$, the corresponding bound state is a BOC. For a BIC to exist when the frequency $\Omega$ is inside the band, an additional condition must be met
\begin{equation}\label{eq-bd-cond-2}
	J(\Omega)=0.
\end{equation}
As shown in Fig.~\ref{fig-Jw}, despite the divergence of the density of states \(\varrho(\omega)\) at the band center, the spectral density \(J(\omega)\) vanishes at the band center (\(\Delta\omega = 0\)) due to the coupling structure of \(V_{\vec{k}}\) given in Eq.~\eqref{eq-Vk}. This satisfies the second criterion for a BIC, as stated in Eq.~\eqref{eq-bd-cond-2}. In what follows, we will further demonstrate that the first criterion, Eq.~\eqref{eq-bd-cond-1}, is also satisfied. Therefore, this point corresponds to a bound state in the continuum (BIC).
Such a BIC does not appear in the 1D infinite CCA case, where the band peaks at the center~\cite{Xiong17}, nor in the 2D CCA systems involving emitters, where the band diverges at the center~\cite{Tudela17L,Tudela17A}. In the following, we will show how a BIC can give rise to intriguing dynamical behavior when the system is resonant with the reservoir.

The cavity field amplitude and the photon number of the system during time evolution can be directly expressed in terms of the two Green's functions~\cite{Xiong10,Zhang12}:
\begin{eqnarray}\label{nt}
  \langle a(t) \rangle &=& u(t) \langle a(0) \rangle,\nonumber\\
  n(t)&=&|u(t)|^2n(0)+v(t),
\end{eqnarray}
where \( n(0) \) denotes the initial photon number of the system. These quantities serve as key physical observables at zero and finite reservoir temperatures, respectively. In the following, we concentrate on analyzing their behavior.

\begin{figure}[ptb]
\centering
\includegraphics[width=8cm]{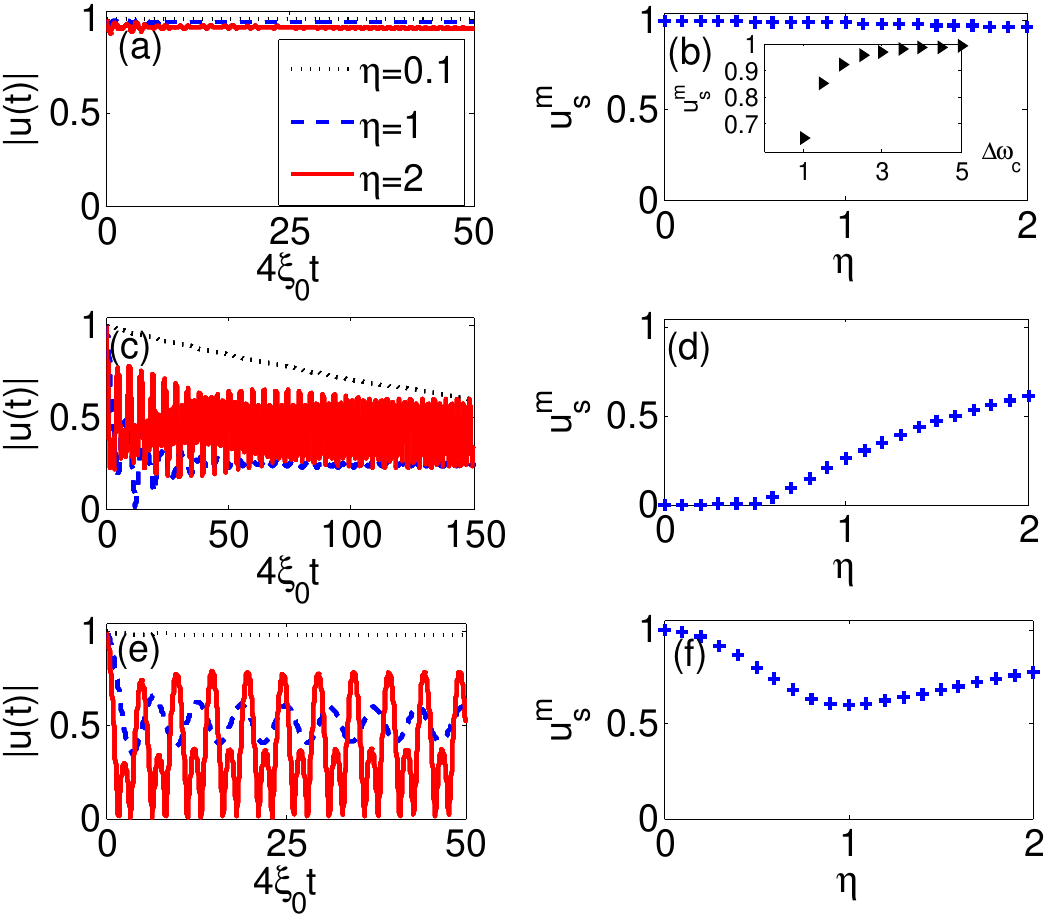}\newline
\caption{Exact dynamics of $|u(t)|$ and its maximum steady value $u^m_s$ for different detunings and coupling strengths. For Figs. (a) and (b), the detuning $\Delta\omega_c=5$. For Figs. (c) and (d), $\Delta\omega_c=0.5$. For Figs. (e) and (f), $\Delta\omega_c=0$. The inset in Fig. (b) shows the maximum steady value $u^m_s$ versus the detuning outside the band when the coupling strength $\eta=1$. In this work, the physical parameters for numerical calculation are the same as the 1D case \cite{Wu10}: the frequency of the cavities in the reservoir is $\omega_0=12.15\text{GHz}=50.25\mu \text{eV}$, and the coupling between the nearest-neighboring cavities in the reservoir is $\xi_0=1.24\mu \text{eV}$.}
\label{fig-ut}
\end{figure}

\subsection{Dynamics at the zero reservoir temperature}
\begin{figure}[ptb]
\centering
\includegraphics[width=8cm]{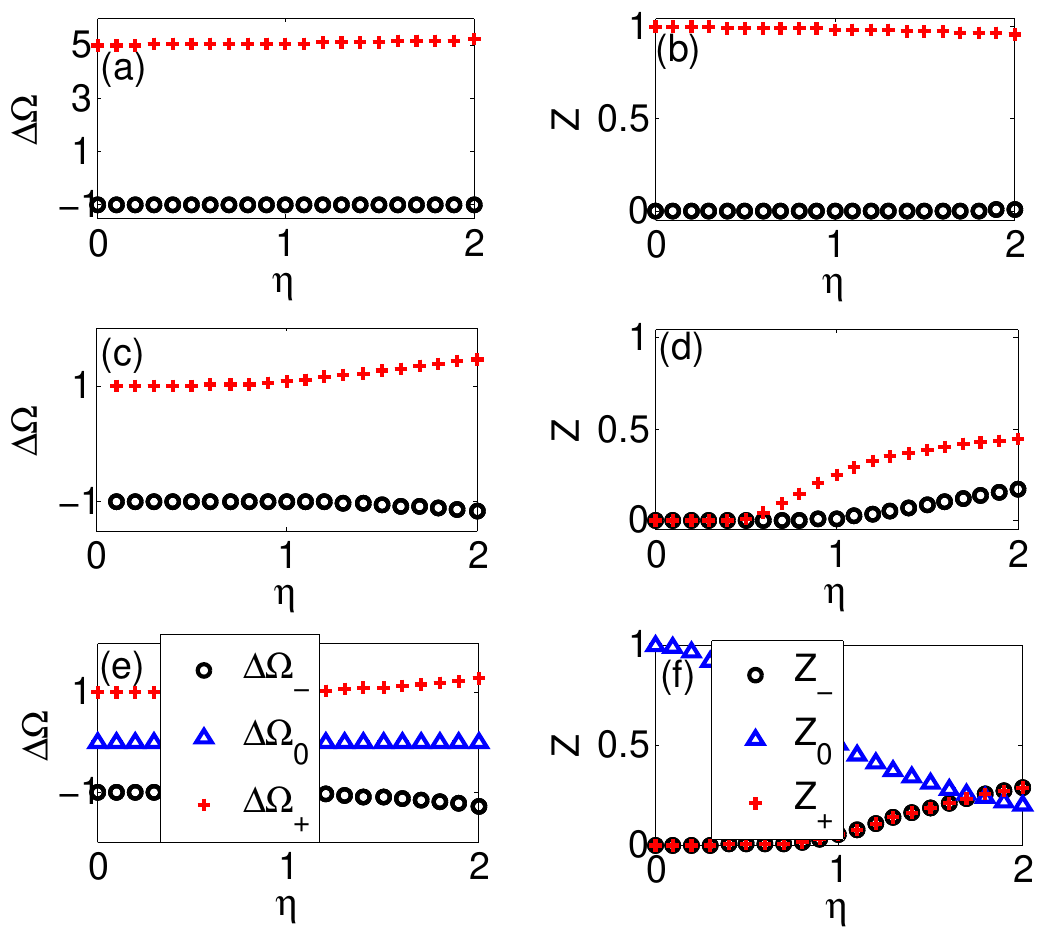}\newline
\caption{(a) Distribution of the eigenenergies $\{\Delta\Omega_j \}$ $(j=\pm,0)$ of the bound states and (b) the corresponding amplitudes $\{Z_j \}$ versus the coupling strength $\eta$. For Figs. (a) and (b), the detuning $\Delta\omega_c=5$. For Figs. (c) and (d), $\Delta\omega_c=0.5$. For Figs. (e) and (f), $\Delta\omega_c=0$. }
\label{fig-wB}
\end{figure}

From Eq.~\eqref{nt}, the absolute value of \( u(t) \) characterizes the dynamic amplification of both the field amplitude and the photon number when the reservoir is initially at absolute zero temperature.
Figure~\ref{fig-ut} shows numerical results for \(|u(t)|\) as a function of the detuning \(\Delta\omega_c \equiv \frac{\omega_c - \omega_0}{4\xi_0}\) and the coupling strength \(\eta\). Here, we extract the maximum value of \(|u(t)|\) attained by the system after it reaches its steady state at time \( t_s \) (or in the long-time limit), denoted as \( u_s^m \equiv \max_{t \geq t_s} |u(t)| \).
Analytically, the solution for \( u(t) \) can be expressed in the form~\cite{Zhang12,Xiong15,Xiong22} (see Appendix~\ref{Appendix-ut}):

\begin{eqnarray}\label{eq-analy-ut}
u(t)=\sum_j {Z}_{j} e^{-i\Omega_{j}t}+\int \frac{d\omega}{2\pi} \mathcal{D}_c(\omega)e^{-i\omega t},\label{eq-ut}
\end{eqnarray}
where ${Z}_{j}$ is the amplitude of the dissipationless steady dynamics of $u(t)$ arisen from the eigenenergy $\Omega_{j}$ of the bound states. And \begin{eqnarray}
\mathcal{D}_c(\omega)=\frac{J(\omega)}{[\omega-\omega_c-\Delta(\omega)]^2+J^2(\omega)/4} \label{eq-Dc}
\end{eqnarray}
is the dissipation spectrum of the continuous energy, which leads to the  (nonexponential) decay of the second term of $u(t)$. As a result, only the bound states contribute to the steady dynamics of the system
\begin{eqnarray}
  u(t_s) = \sum_j {Z}_{j} e^{-i\Omega_{j}t_s}.\label{eq-uts}
\end{eqnarray}
Here the bound states satisfy the condition of (\ref{eq-bd-cond-1}).
Corresponding to Fig. \ref{fig-ut}, Fig. \ref{fig-wB} shows the position of the eigen-energies $\Delta\Omega_{j}\equiv \frac{\Omega_{j}-\omega_0}{4\xi_0}$ ($j=\pm,0$) and the relevant amplitudes $Z_{j}$ for different $\Delta\omega_c$ and $\eta$. Let's combine them to analyze the results.

First, we consider the case of the system frequency $\omega_c$ lies in the band gaps, i.e., $|\Delta\omega_c|>1$. As shown in Figs. \ref{fig-ut}(a) and (b), during time evolution, $|u(t)|$ remains close to its maximum $1$ (or exhibits very small oscillations for large $\eta$) , regardless of the increase in $\eta$. This is the typical property of band gaps, stemming from the effective system-reservoir decoupling as in the 1D case \cite{Xiong17,Wu10}. On the other hand, from Figs. \ref{fig-wB}(a) and (b), it's evident that only one BOC substantially contributes to the system dynamics. Consequently, from Eq.~(\ref{eq-uts}), we have $u(t_s)\approx Z_b e^{-i\Omega_b t_s}$ where $\Omega_b=\Omega_+$ or $\Omega_-$. Thus,
\begin{equation}
|u({t_s})|\approx |Z_b|. \label{eq-uts-1}
\end{equation}
Explicitly, when $\Delta\omega_c=5$, the eigenenergy of the BOC is also approximately $\Delta\Omega_+=5$, with the corresponding amplitude $Z_+\approx 1$. Thus $|u({t_s})|\approx |Z_+| \approx 1$. Notably, although this result resembles the Born-Markovian limit, where $u_{\text{BM}}(t)=e^{-i(\omega'_a+\kappa)t}$ as in Appendix \ref{Appendix-BM}.
For large $\eta$, figure \ref{fig-ut} (a) exhibits slight oscillations in the dynamics, especially over short time scales. This oscillation behavior is actually a manifestation of non-Markovian effects. From Figs. \ref{fig-wB} (a) and (b), we observe that as $\eta$ increases, the BOC with eigenenergy $\Omega_-$ begins to emerge alongside the existing BOC with $\Omega_+$. These two BOCs are asymmetrically distributed. In this case, $u(t_s)=Z_+ e^{-i\Omega_+ t}+Z_- e^{-i\Omega_- t}$ and
\begin{equation}
|u(t_s)|=\sqrt{|Z_+|^2+|Z_-|^2+2Z_+ Z_- \cos[(\Omega_+-\Omega_-)t]}. \label{eq-uts-2}
\end{equation}
However, since $Z_-\ll 1$, the oscillation of $|u(t_s)|$ is extremely small. Moreover, the inset in Fig. \ref{fig-ut}(b) shows the maximum steady value $u^m_s$ versus the detuning outside the band, given $\eta=1$. It is observed that $u^m_s$ approaches 1 only for large detunings. The closer it is to the band edge, the smaller $u^m_s$ is.

When \(\omega_c\) lies within the band but away from the center, i.e., \(0 < |\Delta\omega_c| < 1\), as shown in Figs.~\ref{fig-ut}(c) and (d), for weak coupling strength (e.g., \(\eta = 0.1\)), \(|u(t)|\) decays smoothly to zero as \(t \to \infty\), indicating a near-complete loss of the initial information (field amplitude and photon number). This behavior corresponds to the well-known Markovian or Markovian-like dynamics~\cite{Xiong15}. However, as \(\eta\) increases, the strong non-Markovian memory feedback enables the system to periodically retain its initial information. The larger the value of \(\eta\), the more information is preserved, as illustrated in Fig.~\ref{fig-ut}(d).

From Figs.~\ref{fig-wB}(c) and (d), we observe that in this regime, two bound states outside the continuum (BOCs) with frequencies \(\Omega_{\pm}\) symmetrically positioned around the band center may exist. For weak coupling (\(\eta \lesssim 0.5\)), the corresponding amplitudes \(Z_{\pm} = 0\), leading to \(u(t_s) = 0\). For stronger coupling (\(\eta > 0.5\)), \(Z_+\) and \(Z_-\) grow unequally as \(\eta\) increases, resulting in oscillatory steady-state dynamics governed by Eq.~\eqref{eq-uts-2}. This strong non-Markovian effect has been extensively studied in the 1D case~\cite{Wu10,Xiong22}.

Interestingly, in the 2D case, when \(\omega_c\) is exactly at the band center, i.e., \(\Delta\omega_c = 0\), the system exhibits fundamentally different dynamics. As depicted in Figs.~\ref{fig-ut}(e) and (f), for small \(\eta\), the system nearly decouples from the reservoir, causing \(|u(t)|\) to remain close to unity. However, as \(\eta\) increases, the maximum steady-state value \(u_s^m\) initially decreases before rising again, contrasting sharply with the previous cases.

Moreover, from Figs.~\ref{fig-wB}(e) and (f), we note the presence of an additional bound state with frequency \(\Omega_0\) at the band center, alongside the two BOCs at \(\Omega_{\pm}\). This clearly corresponds to a bound state in the continuum (BIC) satisfying conditions~\eqref{eq-bd-cond-1} and~\eqref{eq-bd-cond-2}. Consequently, in this case,
\[
u(t_s) = Z_0 e^{-i \Omega_0 t} + Z_+ e^{-i \Omega_+ t} + Z_- e^{-i \Omega_- t},
\]
so that one has
\begin{widetext}
\begin{equation}
|u(t_s)|=\sqrt{|Z_0|^2+|Z_+|^2+|Z_-|^2+2Z_0 Z_+ \cos[(\Omega_0-\Omega_+)t]+2Z_0 Z_- \cos[(\Omega_0-\Omega_-)t]+2Z_+ Z_- \cos[(\Omega_+-\Omega_-)t]}. \label{eq-uts-3}
\end{equation}
\end{widetext}
For small coupling strength \(\eta\) (\(\eta < 1\)), we have \(Z_{\pm} = 0\), so that \(|u(t_s)| = Z_0\). This indicates that the system dynamics is solely governed by the BIC. Moreover, the smaller \(\eta\) is, the larger the amplitude \(Z_0\) becomes. However, as \(\eta\) increases (\(\eta > 1\)), the amplitudes \(Z_+\) and \(Z_-\) also grow, while \(Z_0\) decreases. This suggests that both the BOCs and the BIC contribute to \(|u(t_s)|\), as evident from Eq.~\eqref{eq-uts-3}. Here, the two BOCs are symmetrically located in the band gaps and have equal amplitudes, \(Z_+ = Z_-\).
As \(\eta\) further increases, the BIC competes with the two BOCs. For sufficiently large \(\eta\), the BOCs dominate the system dynamics. In the steady state, the system may exhibit multiple oscillation frequencies given by \(|\Omega_+ - \Omega_-|\) and \(|\Omega_0 - \Omega_+|\) (noting that \(|\Omega_0 - \Omega_-| = |\Omega_0 - \Omega_+|\)). Furthermore, for very large \(\eta\), \(Z_0 \to 0\), and the system dynamics is primarily determined by the interplay between the two BOCs, resembling the case \(0 < |\Delta\omega_c| < 1\).
In summary, when \(\Delta\omega_c = 0\), the BIC enables nearly perfect (complete and persistent) quantum information storage for small \(\eta\). For large \(\eta\), high-fidelity quantum information preservation occurs periodically due to the competition between the BIC and the BOCs. At extremely large \(\eta\), only the two BOCs contribute to periodic information storage.\\

To summarize, we have investigated the exact dynamics of the system coupled to a zero-temperature reservoir for various frequency detunings and coupling strengths. When the system frequency lies within the band gaps, nearly perfect information preservation is achievable at any coupling strength due to the predominance of a single BOC. In this regime, the system exhibits Markovian or weakly non-Markovian dynamics. When the system frequency is inside the band but not at the center, the presence of two BOCs leads to strong non-Markovian dynamics, resulting in periodic yet imperfect quantum information storage. Furthermore, when the system frequency is tuned exactly to the band center (the resonant case), perfect quantum information storage occurs in the weak coupling regime due to the sole presence of a BIC, exhibiting Markovian or weakly non-Markovian behavior. In the strong coupling regime, periodic information storage arises from the interplay of the BIC and BOCs. Essentially, the BIC can fully replace the band gaps in ensuring perfect quantum information preservation under Markovian or weakly non-Markovian dynamics.
 Additionally, by comparing the dynamics in Fig.~\ref{fig-ut}(e) with those in Figs.~\ref{fig-ut}(a) and (c), we observe that the presence of the BIC enables the system to reach its steady state more rapidly and exhibit slower oscillations in the steady state compared to cases where only BOCs are involved. Thus, when the system frequency resonates with the reservoir, the BIC accelerates stabilization and moderates oscillations, enhancing the controllability of quantum information in experiments.

\vspace{1em}

Finally, we mention that our results for the zero-temperature reservoir can be cross-checked using the one-particle approach discussed in Sec.~IV. In the following subsection, we proceed to analyze the effects of a finite-temperature reservoir on the system dynamics.
\subsection{Dynamics at a finite reservoir temperature}

\begin{figure}[ptb]
\centering
\includegraphics[width=8cm]{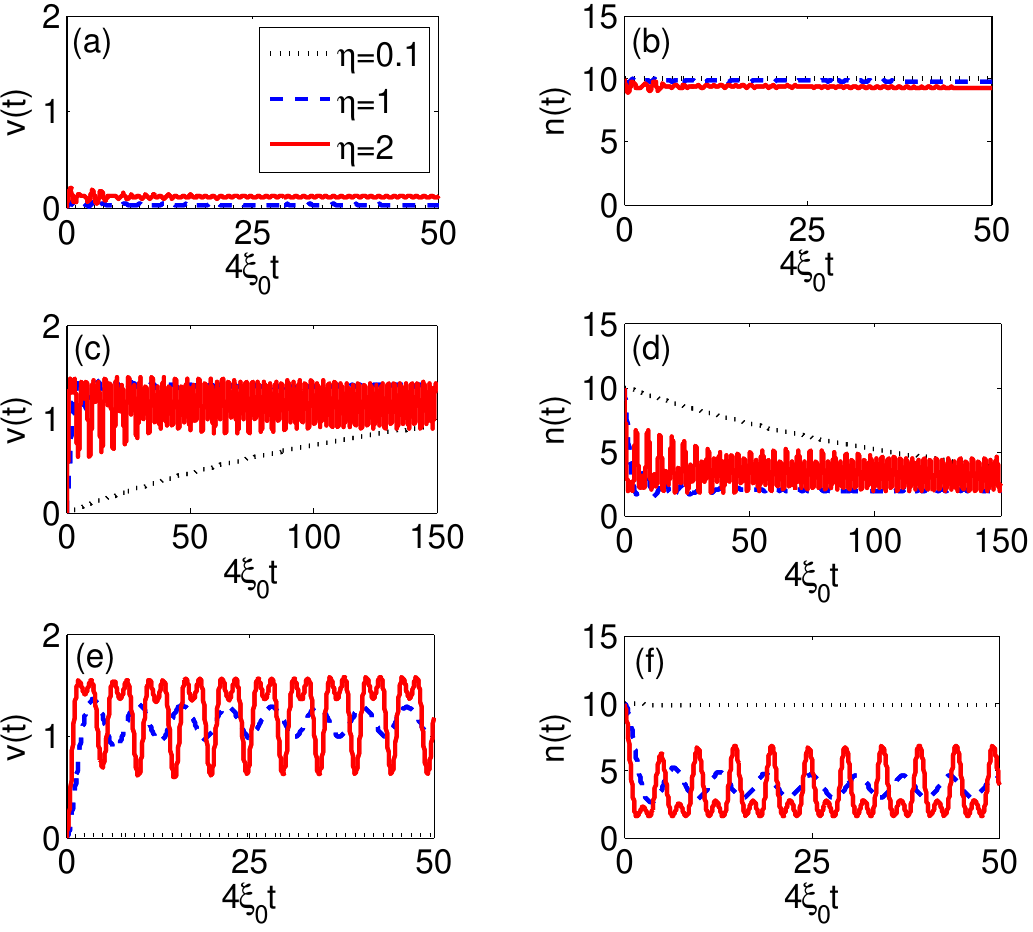}\newline
\caption{Exact dynamics of $v(t)$ and $n(t)$ for different detunings and coupling strengths. Panels (a) and (b) correspond to detuning $\Delta\omega_c=5$, (c) and (d) to $\Delta\omega_c=0.5$, and (e) and (f) to $\Delta\omega_c=0$. Here, $T=2\omega_c$ and the initial photon number is set to $n(0)=10$.}
\label{fig-vt}
\end{figure}

We now examine the dynamics of the Green's function \(v(t)\), which reflects the system's fluctuations, and the photon number \(n(t)\) when the reservoir is at finite temperature, as shown in Fig.~\ref{fig-vt}.

When \(|\Delta\omega_c| > 1\), Figs.~\ref{fig-vt}(a) and (b) show that \(v(t)\) remains extremely small, approaching zero, while the photon number \(n(t)\) stays nearly constant, independent of the coupling strength \(\eta\). Analytically, when only one BOC contributes to \(u(t)\), the steady-state value of \(v(t)\) is given by \cite{Lo15,Xiong15,Zhang19}
\begin{equation}
v(t_s) = \int^{\omega_0+4\xi_0}_{\omega_0-4\xi_0} \frac{d\omega}{2\pi} [\mathcal{D}_b(\omega) + \mathcal{D}_c(\omega)] \bar{n}(\omega,T), \label{eq-vts-1}
\end{equation}
where
\begin{equation}
\mathcal{D}_b(\omega) = \frac{Z_b^2 J(\omega)}{(\omega - \Omega_b)^2}
\end{equation}
represents the contribution from the BOC with frequency \(\Omega_b = \Omega_+\) or \(\Omega_-\).

In the large detuning limit \(|\Delta\omega_c| \gg 1\), both \(\omega_c\) and \(\Omega_b\) lie far from the band edges. Consequently, the denominators of \(\mathcal{D}_b(\omega)\) and \(\mathcal{D}_c(\omega)\) become very large, causing these functions to be negligible and thus \(v(t_s) \to 0\). In the weak coupling regime, this aligns with the Born-Markovian limit where \(v_{\text{BM}}(t) \simeq 0\) (see Appendix~\ref{BM-decouple}). In other words, when the system frequency is within the band gaps, thermal fluctuations are almost absent. According to Eq.~(\ref{nt}), the photon number is then primarily determined by \(u(t)\):
\begin{equation}
n(t_s) \approx |u(t_s)|^2 n(0), \label{eq-nts-1}
\end{equation}
which, for \(|u(t_s)| \approx 1\), implies \(n(t_s) \approx n(0)\). This matches the Born-Markovian result in Eq.~(\ref{BM-decouple}) and is typical of band-gap physics \cite{Wu10}.

Notably, as \(\eta\) increases, oscillations in both \(v(t)\) and \(n(t)\) become more pronounced, especially at short times, as seen in Figs.~\ref{fig-vt}(a) and (b). These oscillations signify non-Markovian effects arising from the simultaneous contributions of two BOCs, where \(\mathcal{D}_b(\omega)\) in Eq.~(\ref{eq-uts-1}) is replaced by \cite{Xiong15}
\begin{equation}
\mathcal{D}_b(\omega) = J(\omega) \frac{Z_+ Z_- \cos[(\Omega_+ - \Omega_-) t]}{(\omega - \Omega_+)(\omega - \Omega_-)}. \label{eq-Dbw-2}
\end{equation}
Additionally, the reservoir's average photon number in our calculations is \(\bar{n}(\omega_c,T) \simeq 1.54\). The behaviors in Fig.~\ref{fig-vt}(a) indicate the system does not settle into thermal equilibrium with the reservoir.

For \(0 < |\Delta\omega_c| < 1\), shown in Figs.~\ref{fig-vt}(c) and (d), \(v(t)\) asymptotically approaches \(\bar{n}(\omega_c,T) \simeq 1.54\). In the weak coupling limit, the steady-state value converges to the Born-Markovian form \(v_{\text{BM}}(t_s) = \bar{n}(\omega_c,T)(1 - e^{-2\kappa t_s})\) as indicated in Eq.~(\ref{BM-utvt}), representing thermal equilibrium between system and reservoir. Consequently, the photon number approaches \(n(t_s) \to \bar{n}(\omega_c,T)\) at long times.

However, with increasing \(\eta\), this thermal equilibrium is disrupted and the steady-state \(v(t_s)\) becomes oscillatory, signaling strong non-Markovian effects. The analytical form of \(v(t_s)\) still follows Eq.~(\ref{eq-vts-1}) but with \(\mathcal{D}_b(\omega)\) given by Eq.~(\ref{eq-Dbw-2}). The photon number is influenced by both \(u(t)\) and \(v(t)\):
\begin{equation}
n(t_s) = |u(t_s)|^2 n(0) + v(t_s). \label{eq-nts-2}
\end{equation}
Stronger coupling \(\eta\) leads to more pronounced oscillations in \(v(t_s)\) and \(n(t_s)\), confirming the non-Markovian nonequilibrium nature of the dynamics.

\begin{figure}[ptb]
\centering
\includegraphics[width=8cm]{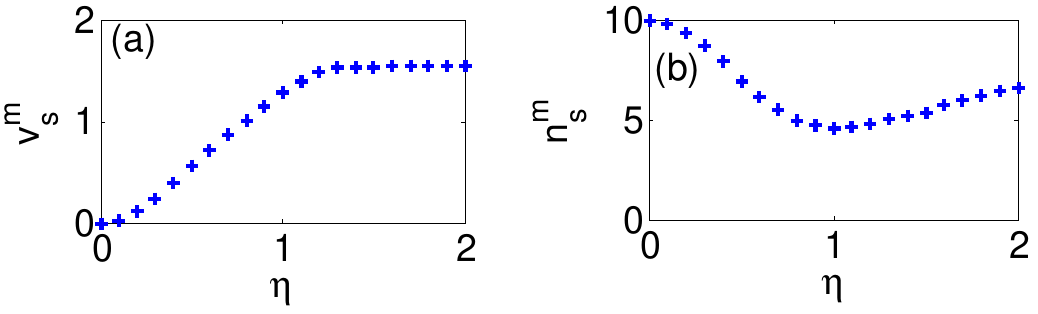}\newline
\caption{Maximum steady-state values of \(v(t)\) and \(n(t)\), denoted \(v^m_s\) and \(n^m_s\), respectively, for varying coupling strength \(\eta\) at \(\Delta\omega_c=0\). Parameters are \(T=2\omega_c\) and \(n(0)=10\).}
\label{fig-vms}
\end{figure}

Finally, at the band center \(\Delta\omega_c = 0\), as illustrated in Figs.~\ref{fig-vt}(e) and (f), the weak coupling regime exhibits \(v(t_s) \simeq 0\) and \(n(t_s) \simeq n(0)\), similar to the large detuning case. Here, \(v(t_s)\) still obeys Eq.~(\ref{eq-vts-1}) but with the BIC contribution
\begin{equation}
\mathcal{D}_b(\omega) = \mathcal{P} \frac{Z_0^2 J(\omega)}{(\omega - \omega_0)^2}, \label{eq-Dbw-3}
\end{equation}
where \(\mathcal{P}\) denotes the Cauchy principal value. Although the denominators here are smaller than in the large-detuning case, the weak coupling makes \(J(\omega)\) sufficiently small, keeping the integral in \(v(t_s)\) minimal and the photon number close to its initial value.

As \(\eta\) grows, \(v(t)\) and \(n(t)\) exhibit oscillations due to non-Markovian effects. All three bound states contribute, and \(\mathcal{D}_b(\omega)\) generalizes to
\begin{widetext}
\begin{equation}
\mathcal{D}_b(\omega) = J(\omega) \left\{ \mathcal{P} \frac{Z_0 Z_+ \cos[(\Omega_0 - \Omega_+) t]}{(\omega - \Omega_0)(\omega - \Omega_+)} + \mathcal{P} \frac{Z_0 Z_- \cos[(\Omega_0 - \Omega_-) t]}{(\omega - \Omega_0)(\omega - \Omega_-)} + \frac{Z_+ Z_- \cos[(\Omega_+ - \Omega_-) t]}{(\omega - \Omega_+)(\omega - \Omega_-)} \right\}. \label{eq-Dbw-4}
\end{equation}
\end{widetext}

In the strong coupling regime, the interplay between the BIC and two BOCs results in oscillatory behavior in both \(v(t_s)\) and \(n(t_s)\). Figure~\ref{fig-vms} depicts the maximum values \(v^m_s\) and \(n^m_s\) as functions of \(\eta\). We observe \(v^m_s\) increases with \(\eta\) up to \(\eta \approx 1.3\), then saturates near \(1.54\). Conversely, \(n^m_s\) decreases for \(\eta < 1\) but rises again beyond \(\eta > 1\), reflecting competing effects of quantum dissipation and thermal fluctuations in Eq.~(\ref{eq-nts-2}). The system remains in a nonequilibrium state with the reservoir.

\vspace{0.5em}

To summarize: For a finite-temperature reservoir, when the system frequency lies in the band gaps, especially at large detuning and weak coupling, thermal fluctuations are negligible and photon number remains nearly constant, consistent with Markovian predictions. Increasing coupling strength induces small oscillations in fluctuations and photon number, indicating weak non-Markovian behavior. When the system frequency is within the band but away from the center, weak coupling leads to standard thermalization towards reservoir photon number, reflecting thermal equilibrium. Stronger coupling breaks this equilibrium, inducing pronounced non-Markovian oscillations due to competing BOCs. At the band center, the weak coupling regime resembles band-gap behavior dominated by the BIC, while strong coupling results in oscillatory dynamics governed by the BIC and two BOCs.

In essence, the role of bound states in finite-temperature dynamics mirrors the zero-temperature case: the BIC acts similarly to band gaps by suppressing thermal fluctuations, and thermal equilibrium emerges only when the system lies within the band (but not at its center) under weak coupling. Otherwise, the system exhibits nonequilibrium dynamics.

Furthermore, the data in Figs.~\ref{fig-vt}(e) and (f) reveal that at \(\Delta\omega_c=0\), the system reaches steady state faster than in other detuning regimes, with slower oscillations. For sufficiently large \(\eta\), steady oscillations in \(v(t)\) and \(n(t)\) display multiple peaks due to the interplay among the three bound states. These features suggest potential advantages for quantum information processing applications near the band center, even at low temperatures, analogous to the zero-temperature scenario.

%\begin{figure}[ptb]
%\centering
%\includegraphics[width=10cm]{Ft.pdf}\newline
%\caption{Exact dynamics of Fringe visibility $F(t)$ for a Schr\"{o}dinger-cat like state and its maximum steady value (denoted as $F^m_s$) for different coupling strengths when $T=2\omega_0$ for (a) and (b) and $T=0$ for (c) and (d). All the values here are given algorithmically with a base of 10. And we set $\Delta\omega_c=0$ and $\alpha_0=2$.}
%\label{fig-Ft}
%\end{figure}

\subsection{Density matrix of the system}
Here, via the evolution of the system's density matrix, we further illustrate the dissipationless dynamics (trapping effect) induced by the bound states.

\begin{figure*}[ptb]
\centering
\includegraphics[width=12cm]{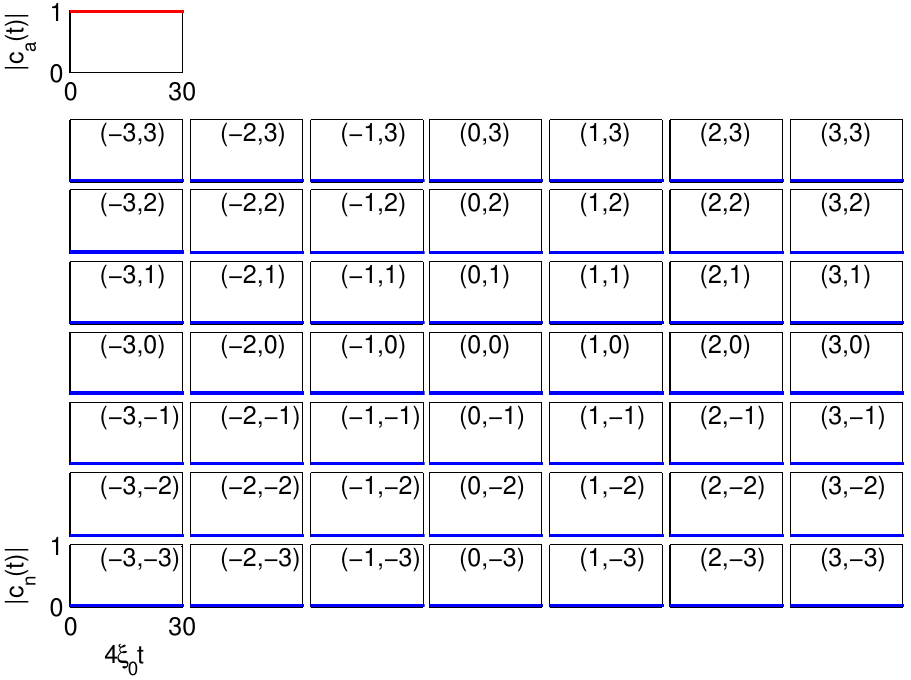}\newline
\caption{Exact dynamics of the amplitudes \(c_a(t)\) (red line) and \(c_{\vec{n}}(t)\) at different lattice sites \(\vec{n}=(x,y)\) (blue lines) for weak coupling \(\eta=0.01\). The indices \(x\) and \(y\) range from \(-3\) to \(3\), and the detuning is set to \(\Delta \omega_c=0\).}
\label{fig-ct-eta01}
\end{figure*}

The exact master equation (\ref{Master equation}) holds for an arbitrary initial state of the target cavity. Consider an initial state given by
\begin{eqnarray}
\rho (0)  =\sum_{m,n=0}^{\infty}c_{mn}|m\rangle\langle n|, \label{rho-0}
\end{eqnarray}
the exact time-evolved state at time $t$ is expressed as \cite{Xiong15}
\begin{equation}
\rho\left(  t\right)  =\!\!\! \sum_{m,n=0}^{\infty}\!\!\!c_{mn}\!\!\!\! \sum_{k=0}^{\min
\{m,n\}}\!\!\!\!\! \!\! d_{k}{\mathcal{A}}_{mk}^{\dagger}\left(  t\right)  \widetilde{\rho
}(v(t))  {\mathcal{A}}_{nk}\left(  t\right)  ,
\label{rho-t}%
\end{equation}
where the kernel $\widetilde{\rho}(v(t))  =\sum_{n=0}^{\infty}\frac{
 [v(t) ]^{n}}{[ 1+v(t) ]  ^{n+1}}|n\rangle\langle
n|$ resembles a thermal state, the raising operator is defined as ${\mathcal{A}}_{mk}^{\dagger}(t)
=\frac{\sqrt{m!}}{( m-k) !\sqrt{k!}}\big[\frac{u(t)}{1+v(t)}
a^{\dagger}\big] ^{m-k}$,  and $d_{k}=[ 1-\frac{|u(t)|
^{2}}{1+v(t)}] ^{k}$.

A special case arises when the Green's functions satisfy
\begin{eqnarray}\label{eq-cond}
  u(t)&\rightarrow  & 1 \nonumber\\
  v(t)&\rightarrow  & 0.
\end{eqnarray}
Under these conditions, the kernel reduces to $\widetilde{\rho}(v(t)) \rightarrow |0 \rangle\langle 0|$, and $d_{k} \rightarrow 0^k$. As a result, only the $k=0$ term contributes to the summation over $k$ in (\ref{rho-t}), yielding
\begin{eqnarray}
\rho\left(  t\right) &\rightarrow& \sum_{m,n=0}^{\infty} c_{mn}  {\mathcal{A}}_{m0}^{\dagger}\left(  t\right)  | 0 \rangle\langle 0 |  {\mathcal{A}}_{n0}\left(  t\right) \nonumber\\
&=& \sum_{m,n=0}^{\infty} c_{mn}  | m \rangle\langle n | .
\label{rho-ts}%
\end{eqnarray}
This indicates that the system returns almost exactly to its initial state. Moreover, all the decay coefficients in (\ref{coefficients}) approach 0. Therefore, if the conditions in (\ref{eq-cond}) are approximately maintained throughout the evolution, the system remains nearly trapped in a localized state and undergoes effectively dissipationless dynamics.

Interestingly, the previous results in Figs. \ref{fig-ut}(a), \ref{fig-ut}(e), \ref{fig-vt}(a) and \ref{fig-vt}(e) indicate that the condition in (\ref{eq-cond}) are nearly satisfied throughout the evolution in two scenarios: (i) when the system frequency lies within a band gap or (ii) when it is at the band center in the weak coupling regime. Notably, these conclusions hold regardless of whether the reservoir is at zero or finite temperature. According to the distribution of the bound states in Fig. \ref{fig-wB}, both cases arise from the emergence of three isolated bound states. In addition to the BOCs in the band gaps, the BIC here results in nearly perfect information trapping within the system, preventing information leakage into the reservoir, irrespective of its temperature. This makes the BIC a promising candidate for quantum memory applications. In fact, a more intuitive depiction of the spatial distribution and localization of the BIC are analyzed in Sec.~IV, where we further demonstrate that the BIC is predominantly localized in the target cavity. Essentially, this dissipationless dynamics governed by the BIC is universal for any initial state of the cavity. As a result, arbitrary quantum states, including single-photon, multi-photon, and superposition states, encoded in the target cavity can be nearly perfectly preserved over time in the weak coupling regime at resonance. This BIC originates from the linear structure of the Hamiltonian (\ref{HT}) and the vanishing of the spectral density (or coupling strength $V_k$) at the band center. Thus embedding qubits in the target cavity does not alter the form of $V_k$, and thus preserves the BIC. This highlights its potential for hybrid quantum systems.

\section{Single-Particle Formalism}

To further validate our results, we employ the single-particle approach introduced in \cite{Longhi2007}. We begin by deriving the dynamics of the system within the single-particle subspace of the total system.

\begin{figure*}[ptb]
\centering
\includegraphics[width=12cm]{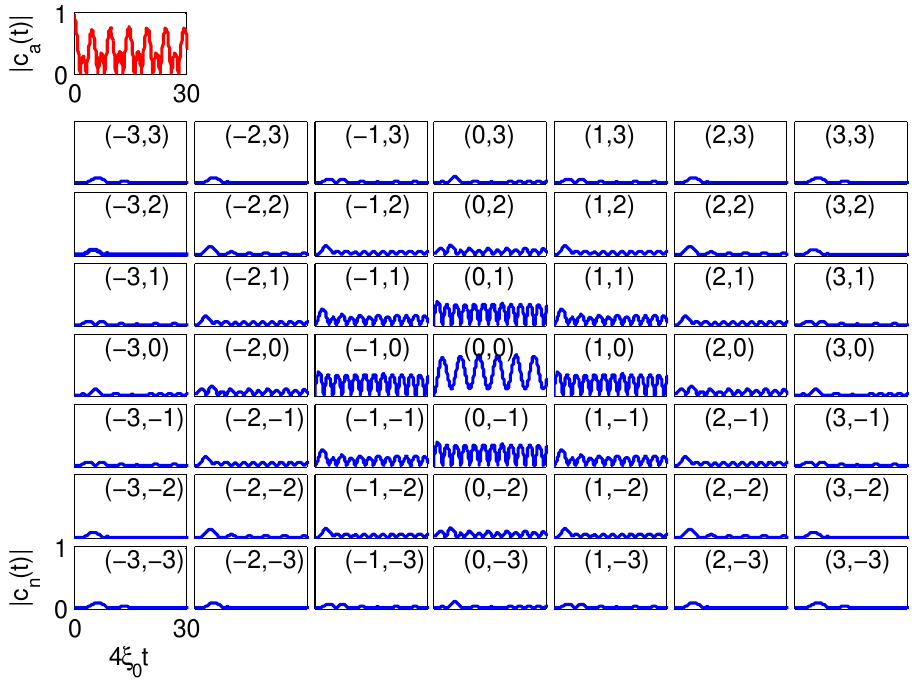}\newline
\caption{Exact dynamics of the amplitudes \(c_a(t)\) (red line) and \(c_{\vec{n}}(t)\) for different sites \(\vec{n} = (x,y)\) (blue lines) under strong coupling \(\eta=2\), with \(x,y \in [-3,3]\) and \(\Delta \omega_c=0\).}
\label{fig-ct-eta2}
\end{figure*}

Starting from the Hamiltonian (\ref{HT}) expressed in coordinate space and assuming there is a single particle (photon) in the entire system, the total state at time \(t\) can be written as
\begin{equation}
|\Psi(t)\rangle = c_a(t) |a\rangle + \sum_{\vec{n}} c_{\vec{n}}(t) |\vec{n}\rangle,
\end{equation}
where \(\vec{n} = (x,y)\) with summations over \(x,y = -N, \ldots, +N\). Here, \(|a\rangle = a^\dagger |\{0\}\rangle\) and \(|\vec{n}\rangle = b_{\vec{n}}^\dagger |\{0\}\rangle\) are single-particle states, and \(|\{0\}\rangle\) is the vacuum state of the total system.
From the Schr\"odinger equation \(i \frac{d}{dt}|\Psi(t)\rangle = H |\Psi(t)\rangle\), the equations of motion for the amplitudes follow:
\begin{align}
\dot{c}_a(t) &= -i \omega_c c_a(t) - i \xi \sum_{\vec{s}} c_{\vec{s}}(t), \label{eq-num-cat}\\
\dot{c}_{\vec{n}}(t) &= -i \omega_0 c_{\vec{n}}(t) + i \xi_0 \sum_{\vec{s}} c_{\vec{n}+\vec{s}}(t) - i \xi c_a(t) \delta_{\vec{n}, \vec{s}}, \label{eq-num-cnt}
\end{align}
where \(\vec{s} = \{(\pm1,0), (0,\pm1)\}\) denotes the four nearest neighbors of the center site \((0,0)\), and \(\delta_{\vec{n}, \vec{s}}\) is the Kronecker delta function.

We solve Eqs.~(\ref{eq-num-cat})--(\ref{eq-num-cnt}) numerically. Figures \ref{fig-ct-eta01} and \ref{fig-ct-eta2} show the amplitude dynamics for weak and strong coupling cases, respectively, with \(\Delta \omega_c = 0\).
In the weak coupling regime (Fig.~\ref{fig-ct-eta01}), the initial excitation localized in the target cavity remains nearly unchanged, as \(|c_a(t)| \approx 1\). This matches the zero-temperature master equation results shown in Fig.~\ref{fig-ut}(a), where \(|u(t)|\) also stays close to unity. The amplitudes \(c_{\vec{n}}(t)\) in the 2D coupled-cavity array (CCA) remain negligible throughout the evolution, indicating minimal excitation transfer from the target cavity. This demonstrates that the bound state in the continuum (BIC) in our model arises from excitation localization in the target cavity with virtually no excitation leakage into the 2D CCA.

This behavior contrasts with the 1D semi-infinite CCA case studied in \cite{Longhi2007}, where the trapping induced by the BIC extends beyond the target system to the connection site (CCA end) and intermediate lattice sites. There, quantum destructive interference suppresses excitation propagation only in the rest of the CCA. Our results instead resemble those for a target system coupled to two semi-infinite CCAs \cite{Ladron06,Voo06}, where the BIC is fully localized in the target system without spreading into the CCAs.

Figure \ref{fig-ct-eta2} illustrates the strong coupling regime. At resonance \(\Delta \omega_c=0\), the amplitude \(|c_a(t)|\) exhibits oscillations analogous to \(|u(t)|\) in Fig.~\ref{fig-ut}(c). The combined effect of the BIC and the bound states outside the continuum (BOCs) leads to steady-state oscillations in the target cavity with two distinct frequencies. Simultaneously, the excitation propagates into the 2D CCA, with cavities closer to the target site exhibiting higher excitation amplitudes in the steady state. As expected, excitation decays with distance, resembling conventional wave propagation.
Interestingly, the cavity at site \((0,0)\), which does not directly couple to the target cavity, attains the highest excitation amplitude within the CCA. This is because its strong indirect coupling via nearest neighbors mediates an effective interaction with the target cavity, resulting in excitation trapping due to the interplay between the BIC and BOCs.

In summary, when the system frequency lies at the center of the 2D CCA energy band, the dynamics differ markedly depending on coupling strength. For weak coupling, the BIC dominates and effectively traps the initial excitation in the target cavity without propagation into the CCA, consistent with the perfect information storage discussed in Sec.~III. For strong coupling, the interaction between the BIC and BOCs causes the excitation to oscillate in the target cavity and simultaneously propagate into the CCA in a wave-like manner.

\section{Conclusions and Outlook}

In this work, we have studied the exact dynamics of a continuous-variable system, a target cavity, embedded in a 2D CCA. Using a master equation approach, we analyzed the system's evolution at both zero and finite reservoir temperatures.
A key feature of the model is the vanishing spectral density at the band center, which gives rise to a BIC when the cavity is resonant with the reservoir, alongside two BOCs induced by the band gaps. This combination leads to dynamics that differ fundamentally from those observed in emitter-based systems in 2D CCAs or continuous-variable systems in 1D settings.
At zero temperature, we find that when the system frequency lies within the band, the competition between bound states results in periodic information preservation under strong coupling, reflecting strong non-Markovian effects. However, the preservation is partial and not persistent. Remarkably, when the system is tuned to the band center, the BIC alone enables complete and persistent information storage even under weak coupling, yielding a transition to Markovian or weakly non-Markovian dynamics. Similar behavior occurs at finite reservoir temperatures. The presence of the BIC leads to negligible thermal fluctuations, an almost constant photon number, and perfect information storage throughout the evolution, even as the system undergoes non-equilibrium dynamics. Thus, in resonance with the band center, the system achieves either perfect or high-fidelity information storage across coupling regimes, and additionally exhibits faster stabilization and longer oscillation periods, features that are advantageous for manipulating continuous-variable quantum systems in information processing tasks. These findings were further supported by a single-particle formalism. In the weak-coupling regime, the BIC leads to excitation trapping in the target cavity, while in strong coupling, excitations propagate through the 2D array like conventional waves.

Beyond these results, the 2D CCA platform explored here offers promising opportunities for topological quantum simulation and a versatile testbed for future developments in quantum technologies. By engineering inter-cavity coupling, for instance, through staggered arrangements or periodic modulations, it is possible to realize nontrivial topological band structures within the reservoir as well as chiral relaxation dynamics \cite{Su79,Li22,Haldane88,Ozawa19,Longhichiral}. In such topological environments, the interplay between BICs, BOCs, and topologically protected edge states could give rise to new forms of robust quantum dynamics, potentially enabling long-lived, fault-tolerant quantum memories and novel schemes for information transfer and entanglement generation. This points toward future work on hybrid topological-reservoir systems and the design of photonic architectures that exploit both non-Markovian memory and topological protection.
As a natural extension of this work, it would be also highly interesting to explore the relaxation dynamics of a continuous-variable system coupled to a \emph{non-Hermitian} bath \cite{NH1,NH2,NH3,NH4,NH5,NH6}. In such settings, the interplay between bound states in the continuum  and bound states outside the continuum could lead to novel relaxation behaviors and richer non-Markovian effects, including phenomena like fractional Zeno dynamics \cite{NH4}. Investigating how BICs and BOCs manifest and influence information storage and decoherence in topological and/or non-Hermitian reservoirs may reveal new mechanisms for controlling dissipation and enhancing quantum memory lifetimes, thereby broadening the scope and applicability of our results in more complex, engineered environments.

One notable potential application of the mechanism studied here is its implementation as a quantum memory \cite{Lvovsky2009,Afzelius2015,Heshami2016,MaL2020,WangYF2023,Lukin2001,Phillips2001,Duan2001,RRR2}, which can be compared with existing approaches for quantum information storage. Our system can operate as a quantum memory through three steps:
(i) Writing: A quantum state of light -- potentially an arbitrary one -- is prepared in the target cavity, with its frequency tuned to the band center ($\omega_c = \omega_0$).
(ii) Storage: Under this resonance condition, the BIC forms automatically, trapping the excitation and preserving the quantum information indefinitely in the weak-coupling limit, or sustaining long-lived oscillations in the strong-coupling regime.
(iii) Reading: To retrieve the stored state, the cavity frequency can be dynamically detuned from the band center, coupling the BIC to the continuum and allowing the state to leak into a designated output channel (e.g., a waveguide coupled to the target cavity).
Compared with established quantum memory protocols \cite{Lvovsky2009,Afzelius2015,Heshami2016,MaL2020,WangYF2023}, including Electromagnetically Induced Transparency (EIT), the Duan-Lukin-Cirac-Zoller (DLCZ) protocol, atomic frequency comb, or Raman memories, our BIC-based approach represents a distinct pathway. Notably, EIT \cite{Lukin2001,Phillips2001} and DLCZ \cite{Duan2001} also exploit dark states, which can be regarded as atomic realizations of a BIC. In contrast, our setup implements the BIC entirely within an integrated photonic lattice using quantum states of light, eliminating the need for complex atomic control and cryogenic conditions. Quantum information is stored directly in the photonic "dark state," enabling preservation of single-photon, multi-photon, or superposition states. This all-light approach \cite{RRR3,RRR4,RRR5} offers the potential advantages of room-temperature operation, chip-scale integration, and scalability. More broadly, the results highlight how engineered BICs can be exploited for controlling non-Markovian dynamics and light-matter interactions in photonic lattices, beyond quantum memory applications.

\bigskip
\appendix

\section{Physical realization of our model}  \label{app-phys-model}
We now qualitatively illustrate the physical realization of our model corresponding to the Hamiltonian (\ref{HT}).

\begin{figure}[ptb]
	\centering
	\includegraphics[width=8cm]{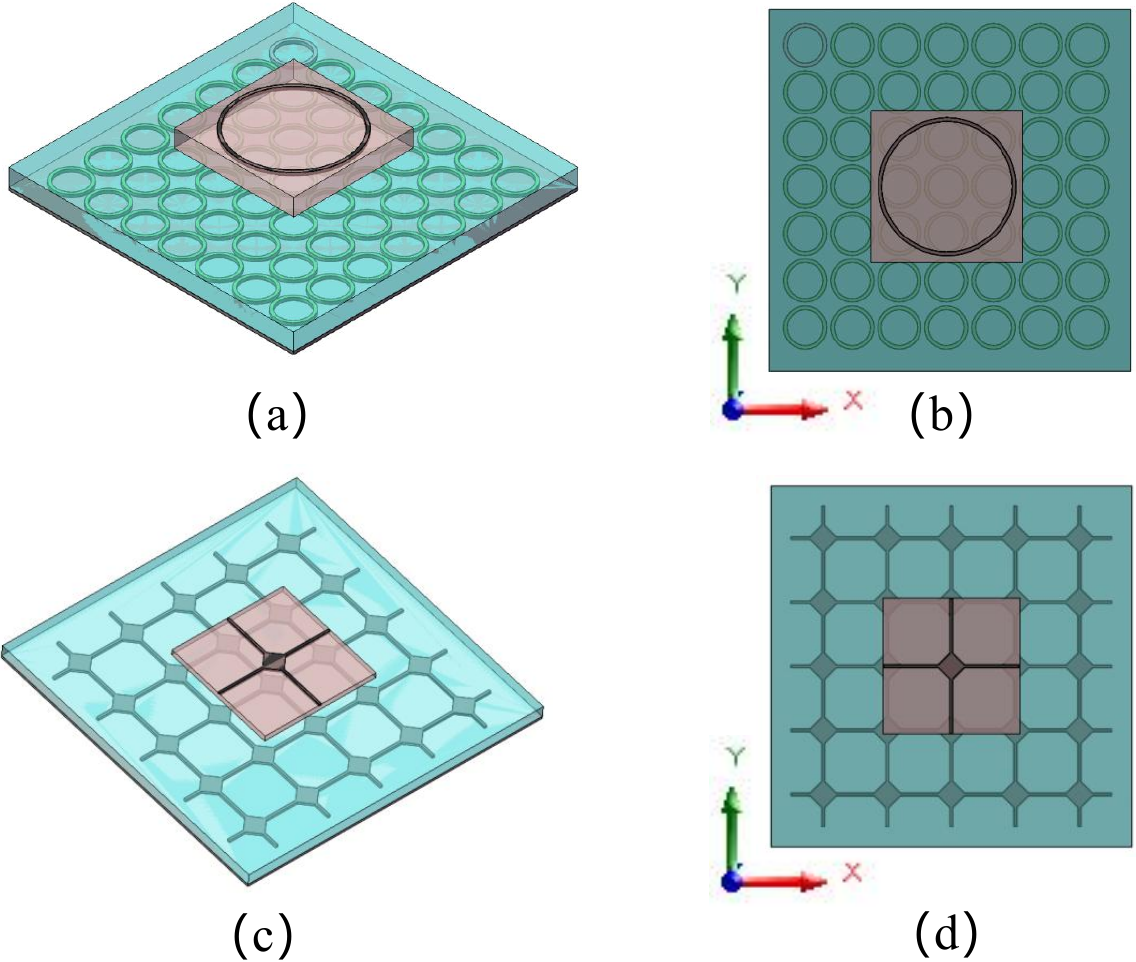}\newline
	\caption{\textbf{Physical realization with ring cavities.} (a) Physical realization of our model with circular cavities (denoted by the circles). The target cavity of the quantum system is the large upper cavity directly above the center of the 2D CCA. (b) The top view projection of (a). (c) Physical realization of our model with square cavities (denoted by the squares). The target cavity is also above the center the 2D CCA. (d) The top view projection of (c).}
	\label{fig-physical-model}
\end{figure}

As shown in Fig. \ref{fig-2D}, the environment of the quantum system is a 2D CCA (Fig. \ref{fig-2D}(a))). Typically, ring cavities provides a good platform for 2D CCAs. It may be of circular or square shapes. Their physical implementations are displayed in Fig. \ref{fig-physical-model}.
These structures are experimentally feasible in various platforms including photonic-crystal cavities \cite{Altug04,Alija06,Gourdon12,Khanikaev17} and semiconductor cavities \cite{Little00,Ohno22,Yang18}.

For the circular cavities in Fig. \ref{fig-physical-model} (a) and (b), the optical field is concentrated near the circumference, enabling direct near-field coupling between neighboring cavities in the 2D CCA. That is, when they are placed closely, their evanescent field overlaps. This corresponds to the coupling term $\sum\limits_{\left\langle\vec{r},\vec{r}'\right\rangle}{{\xi}_{0}}{\left(b_{{\vec{r}}}^{\dagger}{{b}_{{\vec{r}'}}}
+b_{{\vec{r}'}}^{\dagger}{{b}_{{\vec{r}}}}\right)}$ in Hamiltonian (\ref{HT}). The coupling strength $\xi_0$ can be controlled by adjusting the inter-cavity distance and the cavity geometry. On the other hand, the target cavity of the quantum system is a large upper cavity positioned above the center ($(0,0)$) of the 2D CCA. As shown in Fig. \ref{fig-physical-model} (b), there is almost no field overlap between the target cavity and the center cavity directly beneath. Their coupling is negligible. However, the target cavity is almost tangent to the four nearest-neighbor cavities below at $(\pm 1,0)$ and $(0,\pm 1)$, creating significant coupling described by the interaction Hamiltonian $\sum_{\vec{s}}\xi({ b_{{{{\vec{s}}}}}^{\dagger}{a}+a^{\dagger }{{b}_{{{{\vec{s}}}}}} )}$, where ${{\vec{s}}}= \{\left(\pm1,0 \right), \left( 0,\pm 1\right)\}$ denotes their positions in Fig. \ref{fig-2D} (b). In contrast, the target cavity and the four corner cavities at $(\pm{1},\pm{1})$) intersect almost perpendicularly, leading to negligible coupling. This configuration provides a physical realization of the Hamiltonian (\ref{HT}) using circular ring cavities.

For square cavities in Fig. \ref{fig-physical-model} (c) and (d), the coupling between the cavities in the 2D CCA primarily occurs through the four waveguides aligned along the x- and y- directions. The coupling strength $\xi_0$ can thus be controlled by adjusting the geometric parameters of these waveguides.
Additionally, each of the four waveguide terminals in the upper target cavity incorporates a grating that converts horizontal light into vertical light for propagation and reception. This enables the coupling between the upper target cavity and the four nearest-neighbor cavities at ($\{\left(\pm1,0 \right), \left( 0,\pm 1\right)\}$), with the coupling strength $\xi$ being tunable via the gratings.
However, since the waveguide terminals in the lower square cavities lack gratings, negligible coupling exists between the upper target cavity and the center cavity at ($(0,0)$). Furthermore, the next-nearest-neighbor coupling is almost absent between the upper target cavity and the four corner cavities at ($(\pm{1},\pm{1})$). This provides an alternative realization of the Hamiltonian (\ref{HT}).

It should be noted that the above designs omit the coupling between the target cavity and the center cavity at (0,0) to satisfy the form of Hamiltonian (\ref{HT}). However, this omission is not strictly necessary. To generalize, we introduce an additional coupling term as $H_{00} = \xi_{00}(b_{\vec{r}_0}a^\dagger+b_{\vec{r}_0})^\dagger a$, where $\xi_{00}$ the corresponding coupling strength. Adding this term to the interaction Hamiltonian $H_I$ in (\ref{HT}) and performing the Fourier transformation
$b_{\vec{r}}=\iint \frac{d{{\vec{k}}}}{(2\pi)^2}{{e}^{i\vec{k}\cdot \vec{r}}}b_{\vec{k}}$, we obtain an extend  coupling strength:
 \begin{eqnarray}\label{eq-Vkk}
{\tilde{{V}}_{{\vec{k}}}}
 &=&{{e}^{i\vec{k}\cdot {{{\vec{r}}}_{0}}}}[{2\xi }\left( \cos {{k}_{x}}+\cos {{k}_{y}} \right)+\xi_{00}] \nonumber\\
 &=&{{e}^{i\vec{k}\cdot {{{\vec{r}}}_{0}}}}[{\eta }\left( {{\omega }_{0}}-{{\omega }_{{\vec{k}}}} \right)+\xi_{00}]
 \end{eqnarray}
where $\eta\equiv \frac{\xi}{\xi_0}$ as defined in Sec.II. Obviously, the additional coupling $H_{00}$ thus introduces a constant shift in $\tilde{V}_k$, which vanishes when
 \begin{eqnarray}\label{eq-wkk}
 {{\omega }_{{\vec{k}}}}=\omega_0-\frac{\xi_{00}}{\eta}=\omega_0-\frac{\xi_{00}}{\xi}\xi_{0}.
 \end{eqnarray}
 That is, a BIC exists within the band $(\omega_0-4\xi_0,\omega_0+4\xi_0)$ provided that
 \begin{eqnarray}\label{eq-eta00}
 |\frac{\xi_{00}}{\xi}| < 4.
 \end{eqnarray}
This condition implies that the central cavity at (0,0) may couple up to four times more strongly to the target cavity than the four nearest-neighbor cavities at at ($\{\left(\pm1,0 \right), \left( 0,\pm 1\right)\}$). Such a coupling ratio can be feasibly achieved by tuning the geometric distance between the central and target cavities.
Notably, omitting this coupling term in our original model serves to emphasize that the stabilization of the BIC arises predominantly from the four nearest-neighbor interaction topology, which constitutes the fundamental mechanism underlying its formation.

\section{Conditions for bound states}  \label{app-BS}
As shown in Sec. II, the Fourier transformed Hamiltonian (\ref{HT}) in the momentum space is
\begin{eqnarray}\label{H-k}
H&=&{{\omega}_{c}}{a}^{\dagger}{a}+\iint \frac{d{{\vec{k}}}}{(2\pi)^2}~{{{\omega }_{{\vec{k}}}}b_{{\vec{k}}}^{\dagger }{{b}_{{\vec{k}}}}}\nonumber\\
 && +\iint \frac{d{{\vec{k}}}}{(2\pi)^2}~{\left( {{V}_{{\vec{k}}}}b_{{\vec{k}}}^{\dagger }a+V_{{\vec{k}}}^{*}{{b}_{{\vec{k}}}}a^{\dagger } \right)}.
\end{eqnarray}
In terms of the definition of 2D density of states $\varrho\left(\omega\right)=\iint{\frac{d\vec{k}}{{{\left(2\pi\right)}^{2}}}~\delta\left(\omega-{{\omega}_{{\vec{k}}}}\right)}$ , the Hamiltonian (\ref{H-k}) can be reexpressed in the frequency domain
\begin{eqnarray}\label{H-w}
 H&=&{{\omega}_{c}}{a}^{\dagger}{a}+\int d\omega \varrho(\omega)\omega b^{\dagger}_\omega b_\omega \nonumber\\
 && + \int d\omega \varrho(\omega) [ V(\omega)b^\dagger_{\omega}a + V^{*}(\omega)a^\dagger b_{\omega}].
\end{eqnarray}

Since the bound states of the total system are just its eigenstates, we assume the eigenstate corresponding to the eigenenergy (frequency) $\Omega$ is
\begin{eqnarray}\label{Psi-SR}
|\Psi_{SR}\rangle=c_a |a\rangle+\int d\omega \varrho(\omega) c(\omega)|\omega\rangle ,
\end{eqnarray}
where $|a\rangle=a^{\dagger}|\{0\}_{SR}\rangle$ and $|\omega\rangle=b^{\dagger}_{\omega}|\{0\}_{SR}\rangle$ with $|\{0\}_{SR}\rangle$ the vacuum state of the total system. They satisfy the  orthonormality relations as $\langle a|a \rangle=1$, $\langle a|\omega \rangle=0$, and $\omega ' |\omega \rangle=\delta(\omega-\omega ')/\varrho(\omega)$. Of course, the eigenstate $|\Psi_{SR}\rangle$ is normalizable, i.e.,
\begin{eqnarray}
|c_a|^2+\int d\omega \varrho |c(\omega)|^2 < \infty.
\end{eqnarray}
Then from the eigenvalue equation $H|\Psi_{SR}\rangle=\Omega|\Psi_{SR}\rangle$, one can get the following two relations
\begin{eqnarray}
\omega_c c_a +\int d\omega \varrho(\omega) V^{*}(\omega) c(\omega)&=&\Omega c_a, \label{Eig-1} \\
V(\omega) c_a + \omega c(\omega) &=& \Omega c(\omega). \label{Eig-2}
\end{eqnarray}
From (\ref{Eig-2}), one get
\begin{eqnarray}\label{Eig-2.1}
c(\omega)=\frac{V(\omega)}{\Omega-\omega}c_a.
\end{eqnarray}
Substitute it into (\ref{Eig-1}), one obtains
\begin{eqnarray}\label{Eig-3}
\Omega-\omega_c-\Sigma(\Omega)=0
\end{eqnarray}
here $\Sigma(\Omega)=\int \frac{d\omega}{2\pi} \frac{J(\omega)}{\Omega-\omega}$ with $J(\omega)=2\pi\varrho(\omega)|V(\omega)|^2$ the spectral density shown in (\ref{Jw}). (\ref{Eig-3}) is the basic condition for a bound state with frequency $\Omega$. If a bound state with $\Omega$ exists outside the continuum $(\omega_0-4\xi_0,\omega_0+4\xi_0)$, then the bound state is a BOC. While if $\Omega\in(\omega_0-4\xi_0,\omega_0+4\xi_0)$, the corresponding bound state is a BIC if the following additional condition
\begin{eqnarray}\label{Eig-4}
J(\Omega)=0
\end{eqnarray}
is satisfied.
These two conditions (\ref{Eig-3}) and (\ref{Eig-4}) are totally equivalent to the ones obtained in \cite{Miyamoto2005,Longhi2007}.

\section{Analytical solution of $u(t)$ in (\ref{eq-analy-ut})}\label{Appendix-ut}
Below we will give the analytical solution of $u(t)$ in (\ref{eq-analy-ut}) according to Refs. \cite{Zhang12,Zhang19}. The equation of $u(t)$ in (\ref{eq-ut-vt}) takes an intergro-differential form
\begin{eqnarray}
\dot{u}(t)+i{{\omega }_{c}}u(t )+ \int_{{0}}^{t }{\text{d}}{\tau } g(t -{\tau })u({\tau })=0,
\end{eqnarray}
here the integral kernel $f(t)=\int \frac{d\omega}{2\pi} J(\omega) e^{-i\omega t}$ reflects the non-Markovian memory effect of the reservoir on the system.
First, by making a Laplace transformation $U(z)=\int^{\infty}_0 dt~u(t) e^{izt}$, we can get the solution of $u(t)$ in the complex $z$ space as
\begin{eqnarray}
U(z)=\frac{1}{z-\omega_c-\Sigma(z)}
\end{eqnarray}
with $\Sigma(z)$ the reservoir induced self-energy
\begin{eqnarray}
\Sigma(z)= -i\int^{\infty}_0 dt f(t) e^{izt}  =\int \frac{d\omega}{2\pi} \frac{J(\omega)}{z-\omega}
\end{eqnarray}
Then, we can get the solution of $u(t)$ by making the inverse Laplace transformation of $U(z)$
\begin{eqnarray}\label{app-ut}
u(t)=\sum_j {Z}_{j} e^{-i\Omega_{j}t}+\int \frac{d\omega}{2\pi} \mathcal{D}_c(\omega)e^{-i\omega t}.
\end{eqnarray}

The first term of $u(t)$ in (\ref{app-ut}) comes from the poles $\{\Omega_j\}$ of $U(z)$ due to the discontinuity of the spectral density $J(\omega)$.
The poles obey the equation of
\begin{eqnarray}
\Omega_j-\omega_c-\Sigma(\Omega_j)=0,
\end{eqnarray}
which is the same as the equation of the eigenenergies of the bound states in (\ref{Eig-3}). Thus these poles are just the eigenenergies of the total system. As a result, the first term of $u(t)$ in (\ref{app-ut}) gives rise to dissipationless dynamics. Here ${Z}_{j}$ is the corresponding amplitude of the bound state with frequency $\Omega_j$, which equals to the residue of the pole
\begin{eqnarray}
Z_j=\frac{1}{1-\partial_\omega \Sigma(\omega)}|_{\omega=\Omega_j}.
\end{eqnarray}

Conversely, the second term of $u(t)$ in (\ref{app-ut}) gives a dissipation dynamics of the system, which usually exhibits non-exponential decays characterized by the dissipation spectrum
\begin{eqnarray}
\mathcal{D}_c(\omega)=\frac{J(\omega)}{[\omega-\omega_c-\Delta(\omega)]^2+J^2(\omega)/4}. \label{app-Dc}
\end{eqnarray}
Mathematically, this term results from the non-analyticity of $J(\omega)$. When the system reaches its steady state at time $t_s$ or when $t\rightarrow \infty$, the second term will approach zero, and the steady value of $u(t)$ becomes
\begin{eqnarray}
  u(t_s) = \sum_j {Z}_{j} e^{-i\Omega_{j}t_s}.\label{app-uts}
\end{eqnarray}
That is, only the bound states contribute to the steady dynamics of the system.

\section{Born-Markovian limit of the exact master equation (\ref{Master equation})}\label{Appendix-BM}
Here we show the Born-Markovian limit of the exact master equation (\ref{Master equation}). It was already derived in \cite{Xiong10} that in the typical Born-Markovian limit, one can ignore the memory effects of the reservoir on the target cavity by making the perturbation approximation up to the second order of the coupling $V(\omega)$ and taking the long-time Markov limit. Finally, the coefficients (\ref{coefficients}) becomes time independent as follows
\begin{eqnarray}\label{BM-coefficients}
  \omega'_{c} &=& \omega_c+\delta\omega_c, \nonumber\\
  \kappa &=& J(\omega_c)/2, \nonumber\\
  \tilde{\kappa} &=& 2\kappa\bar{n}(\omega_c,T)
\end{eqnarray}
with the frequency shift $\delta\omega_c=\mathcal{P}\int \frac{d\omega}{2\pi} \frac{J(\omega)}{\omega-\omega_c}$
where $\mathcal{P}$ is the principal value of the integral. Equation (\ref{BM-coefficients}) is valid in the weak coupling limit, i.e., $\eta \rightarrow 0$.

In addition, the Born-Markovian limit of the two Green's functions are reduced to
\begin{eqnarray}\label{BM-utvt}
  u_{\text{BM}}(t) &=& e^{-(i\omega'_c+\kappa)t}, \nonumber\\
  v_{\text{BM}}(t) &=& \bar{n}(\omega_c,T)[1-e^{-2\kappa t}].
\end{eqnarray}
And the corresponding physical observable  take the form
\begin{eqnarray}
  \langle a(t) \rangle_{\text{BM}} &=& e^{-(i\omega'_c+\kappa)t}\langle a(0) \rangle, \nonumber\\
  n_{\text{BM}}(t) &=& n(0) e^{-2\kappa t}+  \bar{n}(\omega_c,T)[1-e^{-2\kappa t}].
\end{eqnarray}

In a special case that the decay rate $\kappa=J(\omega_c)/2 = 0$, and then \begin{eqnarray}\label{BM-decouple}
       u_{\text{BM}}(t)& \simeq & e^{-i\omega'_c t}, \nonumber\\
       v_{\text{BM}}(t) & \simeq & 0,\nonumber \\
       n_{\text{BM}}(t) & \simeq & n(0).
     \end{eqnarray}
These three expressions respectively demonstrate that the system undergoes nearly unitary evolution, with zero fluctuation, and the photon number remains unchanged from its initial value. That is, in this special case, the reservoir has no effect on the system, allowing perfect information preservation. For the system investigated in this work, this special case may arise not only when the system frequency $\omega_c$ falls inside the band gaps, but also when $\omega_c$ aligns with the band center.

\bigskip

\textbf{Acknowledgments}
H.~N.~Xiong acknowledges the Fundamental Research Funds for the Provincial Universities of Zhejiang(
Grant No.~RF-A2020002). F. X. Liu acknowledges the National Natural Science Foundation of China (No. 11974015) and the Natural Science Foundation of Zhejiang Province (No. LZ22A040008).

\end{document}